\documentclass[aps,prl,twocolumn,groupedaddress]{revtex4}
\usepackage{graphicx}
\usepackage[fleqn]{amsmath}
\begin{document}

\title{Heat capacity of matter beyond the Dulong-Petit value}

\author{E. I. Andritsos$^{1}$}
\author{E. Zarkadoula$^{1,2}$}
\author{A. E. Phillips$^{1}$}
\author{M. T. Dove$^{1}$}
\author{C. J. Walker$^{1}$}
\author{V. V. Brazhkin$^{3}$}
\author{K. Trachenko$^{1,2}$}
\address{$^1$ School of Physics and Astronomy, Queen Mary University of London, Mile End Road, London, E1 4NS, UK}
\address{$^2$ South East Physics Network}
\address{$^3$ Institute for High Pressure Physics, RAS, 142190, Moscow, Russia}

\begin{abstract}
We propose a new simple way to evaluate the effect of anharmonicity on a system's thermodynamic functions such as heat capacity. In this approach, the contribution of all potentially complicated anharmonic effects to constant-volume heat capacity is evaluated by one parameter only, the coefficient of thermal expansion. Importantly, this approach is applicable not only to crystals but also to glasses and viscous liquids. To support this proposal, we perform molecular dynamics simulations of several crystalline and amorphous solids as well as liquids, and find a good agreement between results from theory and simulations. We observe an interesting non-monotonic behavior of liquid heat capacity with a maximum, and explain this effect as a result of competition between anharmonicity at low temperature and decreasing number of transverse modes at high temperature.
\end{abstract}


\maketitle

\section{Introduction}

One of the central and most recognizable results of statistical physics is the value of constant-volume heat capacity, $C_v$, of a harmonic and classical solid:

\begin{equation}
C_v=3N
\end{equation}
\noindent where $N$ is the number of atoms and $k_{\rm B}=1$. Known as the Dulong-Petit law, Eq. (1) is the result of a solid having $3N$ phonons \cite{landau}.

Experimentally, $C_v$ is almost never $3N$ even in the classical limit $\frac{\hbar\omega_{\rm D}}{T}\ll 1$, where $\omega_{\rm D}$ is Debye frequency, an effect attributed to anharmonicity of interatomic interactions \cite{cowley,marad,anderson,ida,grimvall0,grimbook,grimvall,oga1,dorog,wallace,oga2}. In addition to heat capacity, anharmonicity governs many other properties of condensed matter systems, including thermal expansion, thermal and electric conductivity, elasticity, phase transitions, defect mobility, melting and so on.

There has been a large amount of research into anharmonic effects \cite{cowley,anderson,grimbook} that has resulted in qualitative understanding of the effect anharmonicity on system properties. The common approach is to expand the potential energy $U$ in Taylor series over atomic displacements $u$:

\begin{align}
&U=\frac{1}{2}\sum\limits_{ll^\prime}\phi(r_0^{ll^\prime})+\frac{1}{2}\sum\limits_{ll^\prime x}\phi_x({ll^\prime})(u_x^l-u_x^{l^\prime})+\\ \nonumber
&\frac{1}{4}\sum\limits_{ll^\prime xy}\phi_{xy}(ll^\prime)(u_x^l-u_x^{l^\prime})(u_y^l-u_y^{l^\prime})+\\ \nonumber
&\frac{1}{12}\sum\limits_{ll^\prime xyz}\phi_{xyz}(ll^\prime)(u_x^l-u_x^{l^\prime})(u_y^l-u_y^{l^\prime})(u_z^l-u_z^{l^\prime})+\\ \nonumber
&\frac{1}{48}\sum\limits_{ll^\prime xyz\omega}\phi_{xyz\omega}(ll^\prime)(u_x^l-u_x^{l^\prime})(u_y^l-u_y^{l^\prime})(u_z^l-u_z^{l^\prime})(u_\omega^l-u_\omega^{l^\prime}) \nonumber
\end{align}
\label{expa}
\noindent where the anharmonic coefficients $\phi_{x...}$ are given by the derivatives at equilibrium separations in a usual way \cite{marad}.

As noted by Cowley \cite{cowley}, $\phi_{x...}$ are very complicated to evaluate even if the potential functions are known. Complications related to evaluating $\phi_{x...}$ necessitated approximations, which, as Cowley further notes \cite{cowley}, are quite inadequate for real systems and are useful in order-of-magnitude calculations only. However, assuming that interactions include only pair and short-range (nearest-neighbor) interactions $\phi$ and considering, for example, a face-centered cubic lattice, low-order perturbation theory gives $C_v$ as a function of $\phi$ and $T$ as \cite{marad}:

\begin{align}
C_v=&3N\Big(1-T\frac{1}{8}\frac{\phi^{\rm IV}(r_0)}{(\phi^{''}(r_0))^2}+T\frac{172.3}{4608}\frac{(\phi^{'''}(r_0))^2}{(\phi^{''}(r_0))^3}-\\ \nonumber
&\frac{1}{3}\frac{\hbar^2}{M}\frac{1}{T^2}\phi^{''}(r_0)+O(T^{-3})\Big)
\label{mar}
\end{align}

This relationship is one of the few that provide a closed form for evaluation of $C_v$, assuming that $\phi$ are known and, importantly, represent a faithful representation of interatomic forces.

Unfortunately, the quantitative evaluation of anharmonicity effects has remained a challenge, with the frequent result that the accuracy of leading-order anharmonic perturbation theory is unknown and the magnitude of anharmonic terms is challenging to justify \cite{cowley,marad,grimvall0,wallace,fultz}. Experimental data such as phonon lifetimes and frequency shifts can provide quantitative estimates for anharmonicity effects and anharmonic expansion coefficients in particular, although this involves complications, and limits the predictive power of the theory \cite{cowley}. As noted starting from the early studies \cite{marad,grimvall0}, the main problem with the approach based on expansions such as Eq. (2) and subsequent understanding of anharmonic effects is that good-quality models for interatomic forces are not generally available.

The problem of the anharmonic theory relying on the knowledge of interatomic interaction models has been noted earlier \cite{cowley,marad,wallace}. It has been stated that ``undoubtedly the unsatisfactory nature of these models is the limiting factor in our understanding of many anharmonic properties'', and that if anharmonic calculations are to have any quantitative significance, the realistic models are necessary \cite{cowley}. Consequently, theoretical work on interatomic potentials was stated as an essential future effort at the time \cite{cowley}. We note in passing that despite the progress in materials modeling since that time, the problem remains. Indeed, apart from relatively small number of materials, a negligible fraction of all known ones, it has proved impossible to develop a general recipe for successfully mapping the interatomic interactions onto the sets of empirical functions to be used in expansion such as (2). The problem is particularly acute with modern materials which often have complicated interactions in the form of hydrogen-bonded, polymeric and many-body interactions, magnetic correlations, non-trivial band gap changes with temperature, anisotropy, layered structures, large number of distinct atoms in organic and biological systems and other factors that severely limit the development of high-quality interaction models. Importantly, because the anharmonic effects are believed to be small, a small departure of a potential model from a high quality one renders the partitioning into harmonic and anharmonic parts (2) and subsequent interpretation of anharmonicity meaningless.

Another approach to treat anharmonicity is to invoke Gr\"{u}neisen approximation, where the softening of phonon frequencies $\omega_i$ is quantified by parameters $\gamma_i=-\frac{V}{\omega}\left(\frac{\partial\omega_i}{\partial V}\right)_T$, and discuss the macroscopic equations of state \cite{anderson}. However, $C_v$ was not previously calculated in this approach in the form free of adjustable parameters and suitable for direct numerical evaluations.

In view of persisting difficulties of evaluating anharmonic effects, it is important to have an alternative general method of estimation of anharmonic $C_v$. It is also important to have a method applicable not only to crystalline systems, but also to amorphous solids as well as liquids, systems for which the traditional perturbation approaches are not suitable, as discussed below in more detail.

In the course of studying the problem of glass transition, we have proposed \cite{prb} that the effects of anharmonicity on $C_v$ can be evaluated as

\begin{equation}
C_v=3N(1+\alpha T)
\label{10}
\end{equation}
\noindent where $\alpha$ is the coefficient of thermal expansion.

There is no contradiction in the relationship (\ref{10}) between the constant-volume $C_v$ and thermal expansion, as might be perceived. As discussed below in detail, the relationship is due to the softening of bulk modulus with temperature at constant volume due to intrinsic anharmonicity, an effect that can be related to $\alpha$ in Gr\"{u}neisen approximation.

In Eq. (\ref{10}), all potentially complicated effects of anharmonicity discussed above are evaluated by one parameter, $\alpha$. Importantly, $\alpha$ is not an adjustable parameter, but is fixed by system properties. Another important feature of Eq. (\ref{10}) is that $\alpha$ can be independently measured or calculated in a straightforward way no matter how complicated interactions in a system are. Appealingly simple, Eq. (\ref{10}) provides an important and straightforward way of estimating the effect of anharmonicity on $C_v$. Perhaps not unexpectedly, the simplicity is achieved by making approximations, and this paper is partly devoted to assessing these approximations, a point to which we return below.

Importantly, Eq. (\ref{10}) can be used to evaluate anharmonicity in two important types of condensed matter systems, glasses and liquids, for which calculations based on anharmonic expansions such as (2) do not work. Indeed, the evaluation of the anharmonic terms in Eq. (2) and coefficients $\phi_{x...}$ involves sums over wave vectors $k$ in a crystal \cite{cowley,marad}. On the other hand, $k$ are not defined in amorphous glasses, at least not at large $k$. In liquids, an expansion such as (2) can not be made even in principle because atoms do not oscillate around fixed positions as in solids, which is the starting point of theories based on Eq. (2) and similar ones.

In this paper, we extend our new approach, and address the validity of Eq. (\ref{10}) across a wide range of crystals, glasses and viscous liquids. We perform molecular dynamics (MD) simulations, and find a good agreement between simulation results and Eq. (\ref{10}) in several crystalline and amorphous solids as well viscous liquids in a wide temperature range.

We note that using MD simulation to study Eq. (\ref{10}) has two important advantages over experiments. First, experimental $C_v$ is calculated from the measured $C_p$ as $C_v=C_p-VT\alpha^2 B$, where $B$ is the bulk modulus. There are uncertainties in experimentally determined $\alpha$ and $B$, particularly at high temperature, which implies uncertainty in $C_v$ \cite{anderson}. In the MD simulation, this problem does not originate because simulations can be performed at constant volume. Second, the classical limit $\frac{\hbar\omega_{\rm D}}{T}\ll 1$ giving $C_v=3N$ is not achieved in many experimental systems due to high $\omega_{\rm D}$ \cite{anderson}. Consequently, it is often not clear to what extent the deviation of experimental $C_v$ from $3N$ is due to anharmonicity or quantum effect of phonon excitation. This issue does not originate in our MD simulations, which are classical.

We finally note that when evaluations of anharmonic effects are possible for certain systems, traditional perturbation approaches achieve the accuracy at the level of order-of-magnitude agreement with experiments or simulations (see, e.g., Refs \cite{marad,wallace,grimvall0}). We aim for at least the same level of accuracy in our new general method of evaluating anharmonic effects. The accuracy is determined by certain approximations that are used to derive a simple form of Eq. (\ref{10}). We find that Eq. (\ref{10}) gives correct order-of-magnitude evaluation of anharmonic effects, the result that is considered as best of what can be achieved in the traditional perturbation expansion approximations.

\section{Theory}

We start with the derivation of Eq. (\ref{10}). The free energy of a harmonic solid in the high-temperature approximation is $F=3NT\ln\frac{\hbar\bar{\omega}}{T}$, where $\bar{\omega}^{3N}=\omega_1\omega_2...\omega_{3N}$ is geometrically averaged phonon frequency \cite{landau}. In the harmonic case, $\bar{\omega}$ is constant, giving the entropy $S=-\left(\frac{\partial F}{\partial T}\right)_v=3N\left(1+\ln\frac{T}{\hbar\bar{\omega}}\right)$ and $C_v=T\left(\frac{\partial S}{\partial T}\right)_v=3N$. Anharmonicity results in the decrease of $\bar{\omega}$ with temperature. Then, $S=3N\left(1+\ln\frac{T}{\hbar\bar{\omega}}-\frac{T}{\bar{\omega}}\frac{{\rm d}\bar{\omega}}{{\rm d}T}\right)$, and

\begin{equation}
C_v=3N\left(1-\frac{2T}{\bar{\omega}}\frac{{\rm d}\bar{\omega}}{{\rm d}T}+ \frac{T^2}{\bar{\omega}^2}\left(\frac{{\rm d}\bar{\omega}}{{\rm
d}T}\right)^2-\frac{T^2}{\bar{\omega}}\frac{{\rm d^2}\bar{\omega}}{{\rm d}T^2}\right)
\label{cv}
\end{equation}

\noindent where the derivatives are taken at constant volume.

In the high-temperature limit where $F=3NT\ln\frac{\hbar\bar{\omega}}{T}$, Eq. (\ref{cv}) is exact, and is the starting point of our theory. Evaluation of $C_v$ requires the knowledge of $\frac{{\rm d}\bar{\omega}}{{\rm d}T}$, which we calculate below.

The phonon pressure, $P_{\mathrm{ph}}$, is $P_{\mathrm{ph}}=-\left(\frac{\partial F}{\partial V}\right)_T=\frac{3NT\gamma}{V}$, where $\gamma$ is the average Gr\"{u}neisen parameter $\gamma=\frac{1}{3N}\sum\limits_{i=1}^{3N}\gamma_i$ and $\gamma_i=-\frac{V}{\omega_i}\left(\frac{\partial\omega_i}{\partial V}\right)_T$ \cite{anderson}. This gives the bulk modulus $B_{\mathrm{ph}}=-\frac{3NT\gamma(q-1)}{V}$ and $\left(\frac{\partial B_{\mathrm{ph}}}{\partial T}\right)_v=-\frac{3N\gamma(q-1)}{V}$, where $q=\frac{\partial\ln\gamma}{\partial\ln V}$. Experimentally, $q$ is known to be fairly constant across the range of systems (e.g. $q$=2.1 for Pb, 3.2 for Ge \cite{grimbook}, 1.4 for MgO \cite{anderson}, 1.5--2 for alkali halides \cite{roberts}, 1.7 for MgSiO$_3$ perovskite \cite{stix} etc). For simplicity, we set $\left(\frac{\partial B_{\mathrm{ph}}}{\partial T}\right)_v=-\frac{3N\gamma}{V}$ as this does not affect our order-of-magnitude evaluations of $c_v$, a point to which we return below. Using $\gamma=\frac{V\alpha B}{C_v}$ and $B=B_0+B_{\mathrm{ph}}$, where $B$ and $B_0$ is the total and static bulk modulus, respectively, we find

\begin{equation}
\left(\frac{\partial B_{\mathrm{ph}}}{\partial T}\right)_v=-\alpha(B_0+B_{\mathrm{ph}})
\label{bulk}
\end{equation}

\noindent where we set $C_v=3N$ in this approximation.

For small $\alpha T$, which is often the case in the experimental temperature range, Eq. (\ref{bulk}) implies $B\propto -T$, consistent with the experiments \cite{anderson}. We note that experimentally, $B$ linearly decreases with temperature at both constant pressure and constant volume (constant-volume decrease can be small in some systems) \cite{anderson,gold,and1,yamamo}. The decrease of $B$ with $T$ at constant volume is due to the intrinsic anharmonicity related to the softening of interatomic potential at large vibrational amplitudes; the decrease of $B$ at constant pressure has an additional contribution from thermal expansion.

The next step is to assume that $\bar{\omega}^2\propto B$, a relationship that holds true if $\omega_i^2\propto B$. For acoustic modes, $\omega_i^2\propto B$ because $\omega_i^2=k^2c^2\propto B+\frac{4}{3}G$ and the shear modulus $G$ scales with $B$ via the Poisson ratio that is nearly constant in all systems. Therefore, $\bar{\omega}^2\propto B$ is applicable to any system as long as the phonon spectrum is treated in Debye approximation, as is often the case. In a general case of a spectrum that includes optic modes, the relationship $\bar{\omega}^2\propto B$ can be addressed by studying how $\omega_i$ and $B$ change in response of external parameters such as temperature and pressure. It has been found that $\omega_i^2\propto B$ is the case for optic modes in a wide temperature range, both longitudinal and transverse \cite{aguado}. The increase of $\omega_i$ including acoustic and optic modes is also seen in a wide pressure range, accompanied by the simultaneous increase of $B$ \cite{klug}.

Finally, combining $\bar{\omega}^2\propto B_0+B_{\mathrm{ph}}$ and Eq. (\ref{bulk}), we find $\frac{1}{\bar{\omega}}\left(\frac{{\rm d}\bar{\omega}}{{\rm d}T}\right)_v=-\frac{\alpha}{2}$. Putting the last relationship in Eq. (\ref{cv}) gives Eq. (\ref{10}). We note that the last two terms in Eq. (\ref{cv}) cancel out if
$\left(\frac{{\rm d}\bar{\omega}}{{\rm d}T}\right)_v\propto\bar{\omega}$, as is the case here.

As follows from the previous discussion, the evaluation of $c_v$ can be made more precise if values of $q$ are retained in the calculation. In this case, $\left(\frac{\partial B_{\mathrm{ph}}}{\partial T}\right)_v=-\delta(B_0+B_{\mathrm{ph}})$, where $\delta=\alpha(q-1)$. Combining it with $\bar{\omega}^2\propto B_0+B_{\mathrm{ph}}$ gives $\frac{1}{\bar{\omega}}\left(\frac{{\rm d}\bar{\omega}}{{\rm d}T}\right)_v=-\frac{\delta}{2}$. Using it in Eq. (\ref{cv}) gives

\begin{equation}
C_v=3N(1+\delta T)
\label{11}
\end{equation}

Here, similar to Eq. (\ref{10}), all anharmonic effects are represented by one parameter, $\delta$. This parameter quantifies the decrease of $B$ with temperature at constant volume. Concerned with demonstrating an order-of-magnitude evaluation of $c_v$ using our new approach, we will not pursue Eq. (\ref{11}) further, and concentrate on Eq. (\ref{10}).

\section{Molecular dynamics simulations}

We now discuss our MD simulations. We have aimed for diversity of structures and interactions, and consequently chosen several systems with different symmetry, structure and interatomic potentials: crystalline Ge, NaCl, Al$_2$O$_3$ (corundum), TiO$_2$ (rutile), ZrSiO$_4$, SiO$_2$ glass and a model liquid system. For Ge, we used many-body environment-dependent (``bond-order'') Tersoff potential \cite{tersoff}. Here, the interaction strength between any two atoms depends on their environment and coordination. This potential therefore represents an example of a crystalline system which can not be treated in the expansion approach (2). Empirical potentials for Al$_2$O$_3$ \cite{cor}, TiO$_2$ \cite{tio2}, NaCl \cite{martin} ZrSiO$_4$ \cite{zir1,zir2,zir3} and SiO$_2$ glass \cite{tsu} included long-range Coloumb and short-range Buckingham or Morse interactions. Ref. \cite{sio2} discusses details of generation of SiO$_2$ glass structure. For the liquid, we employed Lennard-Jones (LJ) potentials designed to simulate a binary liquid in the supercooled viscous state \cite{lj}. The binary liquid consists of two distinct atomic types with different interaction parameters and effective sizes to avoid crystallization at low temperature.

We note here that the empirical potentials we employed may or may not closely reproduce the experimental $\alpha$ or other properties such as $c_v$ or $B$. However, this is not important for our study as we aim to show that a given force field, even though approximate, still results in the relationship between $c_v$ and $\alpha$ given by Eq. (\ref{10}). In this sense, it is only important that a force field gives physically sensible set of $\omega_i$ (e.g., real $\omega_i$) and other physical characteristics such as elasticity and thermal expansion that show commonly observed temperature dependence, because our derivation of Eq. (\ref{10}) is based on these properties and relationships between them.

We have used DL\_POLY programme \cite{dlpoly} for our MD simulations. For solids, the number of atoms was in the 12,000-27,000 range depending on the system. For the LJ liquid, we used 64,000 atoms. We have verified that increasing the number of atoms does not change the results. The energy of the system, $E$, was calculated in constant-energy and volume ensemble simulations by equilibrating the system at a given temperature. The system volume and $\alpha$ were calculated in constant-pressure ensemble simulations. We have performed simulations in a wide temperature range (see Figure 1) with temperature step of 1 K. Each temperature point was simulated on a separate processor using our high-throughput computing cluster. The constant-volume specific heat was calculated as $c_v=\frac{1}{N}\frac{{\rm d}E}{{\rm d}T}$. To reduce the fluctuations of the derivative, we have fitted the energy using high-order polynomials and cubic splines, and verified that $c_v$ is not sensitive to the polynomial order used and fitting parameters.

In Figures 1--2 we show the calculated $c_v$ and relative volume $\frac{V}{V_0}$, where $V_0$ is the system volume at the lowest simulated temperature, for 6 solid systems and for the LJ liquid. We observe that $c_v$ for different systems increase above the Dulong-Petit value of 3 in the wide temperature range. We found an exception to this behavior in crystalline Ar, where the increase of $c_v$ is preceded by its decrease at low temperature. In soft crystals such as Ar with large anharmonicity ($\gamma=$3.5), we do not expect the starting assumptions of the Gr\"{u}neisen approximation that we employed here to hold. Figures 1--2 enable us to see how well the slope of $c_v$ predicted by Eq. (\ref{10}) agrees with the actual value of $\alpha$. Consequently, we calculated $\alpha_c$ from Figure 1a as $c_v=3(1+\alpha_c T)$ and $\alpha=\frac{1}{V_0}\frac{\Delta V}{\Delta T}$ from Figure 1b and 2b. For the LJ liquid, $\alpha_c$ was calculated from the linear increase of $c_v$ at low temperature in Fig. 2a, for the reasons discussed below in detail. For some systems, $c_v$ and $\frac{V}{V_0}$ are not linear with temperature in the whole temperature range. In this case, we have calculated $\alpha_c$ and $\alpha$ at each temperature, and have taken the average.

\begin{figure}
\begin{center}
{\scalebox{0.4}{\includegraphics{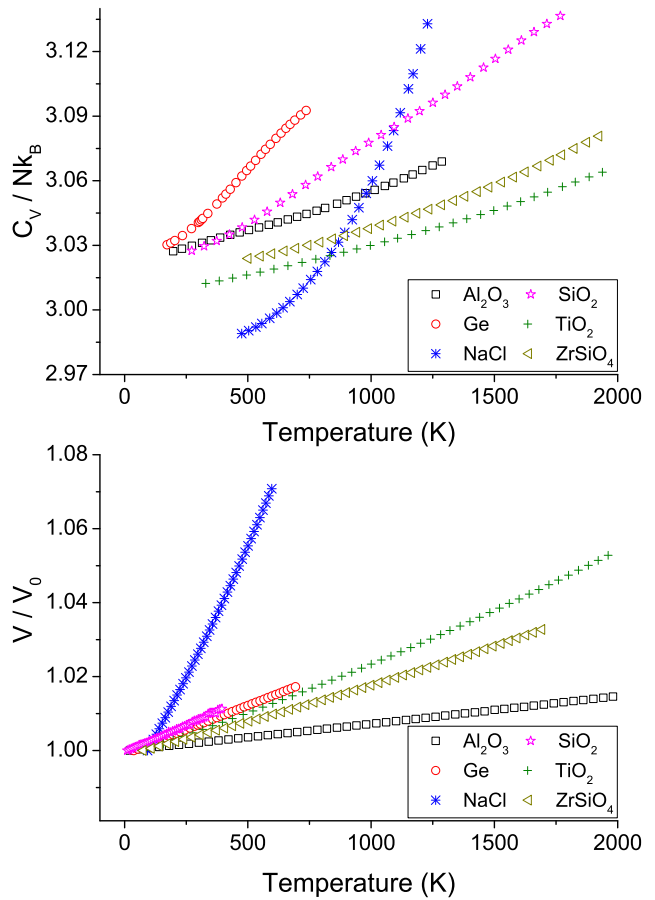}}}
\end{center}
\caption{$c_v$ (a) and $\frac{V}{V_0}$ (b) for simulated crystalline and amorphous systems.}
\label{1}
\end{figure}

\begin{figure}
\begin{center}
{\scalebox{0.4}{\includegraphics{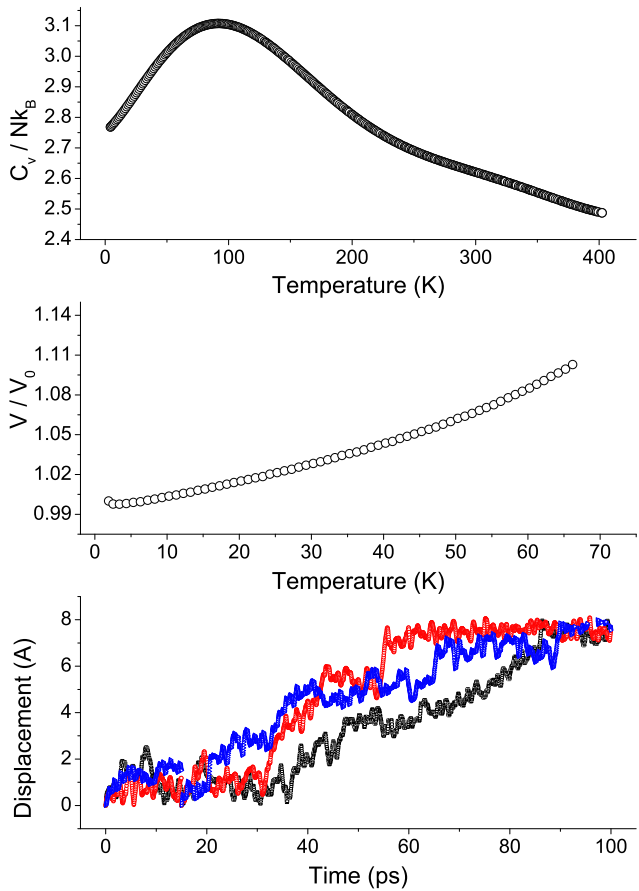}}}
\end{center}
\caption{$c_v$ (a) and $\frac{V}{V_0}$ (b) for LJ liquid. (c) shows coordinates of three atoms with large atomic displacements.}
\label{2}
\end{figure}

The calculated values of $\alpha_c$ and $\alpha$ are: crystalline Ge ($\alpha_c=3.6\cdot 10^{-5}$ K$^{-1}$, $\alpha=2.6\cdot 10^{-5}$ K$^{-1}$), TiO$_2$ ($\alpha_c=1.1\cdot 10^{-5}$ K$^{-1}$, $\alpha=2.8\cdot 10^{-5}$ K$^{-1}$), NaCl ($\alpha_c=7\cdot 10^{-5}$ K$^{-1}$, $\alpha=14\cdot 10^{-5}$ K$^{-1}$), ZrSiO$_4$ ($\alpha_c=1.3\cdot 10^{-5}$ K$^{-1}$, $\alpha=2\cdot 10^{-5}$ K$^{-1}$), Al$_2$O$_3$ ($\alpha_c=1.3\cdot 10^{-5}$ K$^{-1}$, $\alpha=0.7\cdot 10^{-5}$ K$^{-1}$), SiO$_2$ glass ($\alpha_c=2.4\cdot 10^{-5}$ K$^{-1}$, $\alpha=2.9\cdot 10^{-5}$ K$^{-1}$), LJ liquid ($\alpha_c=1.75\cdot 10^{-3}$ K$^{-1}$, $\alpha=1.72\cdot 10^{-3}$ K$^{-1}$).

We observe that the anharmonic contribution to $c_v$ can be evaluated by Eq. (\ref{10}) fairly well. $\frac{\lvert\alpha_c-\alpha\rvert}{\alpha}$, averaged over all systems, is within 40\%. The discrepancy between the predicted and calculated values is within the approximations we made in deriving Eq. (\ref{10}), including our neglecting the dependence of $\gamma$ on volume, which can alter the predicted $\alpha_c$ by up to a factor of 2 (see Eq. (\ref{11}) and the discussion preceding Eq. (\ref{bulk})). We therefore find that Eq. (\ref{10}) gives the correct order-of-magnitude evaluation of anharmonic effects, the result that is considered as best of what can be achieved using the traditional perturbation expansion approximations (see, e.g., Refs \cite{marad,wallace,grimvall0}).

\section{Heat capacity of a liquid in the viscous regime}

We now discuss why and how the above theory applies to viscous liquids in addition to solids. We first demonstrate that the temperature range where $c_v$ linearly increases in Fig. 2a, and where we calculate $\alpha_c$, corresponds to a viscous liquid. Following a somewhat general definition, a viscous liquid is a liquid whose relaxation time $\tau$ is much larger than Debye vibration period of about 0.1 ps: $\tau\gg\tau_{\rm D}$. Here, $\tau$ is the average time between consecutive atomic jumps in a liquid at one point in space \cite{frenkel}. In Fig. 2c we plot the coordinates of three atoms from the simulation of the binary LJ liquid at 50 K, corresponding to temperature in the middle of the linear increase of $c_v$ in Fig. 2a. We observe atomic displacements reaching 8 \AA\ during the time of our simulation, witnessing that large-amplitude diffusive motions are present as is the case for liquids. Second, we estimate $\tau$ as the average time between atomic jumps by definition. Averaged over different atoms, $\tau$ is approximately 15 ps. Therefore, $\tau\gg\tau_{\rm D}$, corresponding to a viscous liquid.

There are two types of motion in a liquid: phonon motion that includes one longitudinal mode and two transverse modes with frequency $\omega>\frac{1}{\tau}$, and diffusional motion \cite{frenkel}. Consequently, the total liquid energy, $E$, is the sum of the phonon energy, $E_{\mathrm{ph}}$, and diffusional energy, $E_{\mathrm{dif}}$: $E=E_{\mathrm{ph}}+E_{\mathrm{dif}}$. $E_{\mathrm{dif}}$ includes both kinetic energy of diffusing atoms and potential energy of their interaction with other atoms. As argued by Frenkel, a particle spends time $\tau$ vibrating in between jumps \cite{frenkel}. The time it takes a particle to jump from one equilibrium position to the next is approximately equal to $\tau_{\rm D}$. Therefore, the probability of a jump is $\rho=\frac{\tau_{\rm D}}{\tau}$. In statistical equilibrium, the number of atoms in the transitory diffusing state is $N_{\mathrm{dif}}=N\rho$, where $N$ is the total number of atoms, giving

\begin{equation}
N_{{\mathrm{dif}}}=N\frac{\tau_{\rm D}}{\tau}
\label{num}
\end{equation}

Eq. (\ref{num}) implies that in a viscous liquid where $\tau\gg\tau_{\rm D}$, the relative number of diffusing atoms at any given moment of time is negligible. Consequently, $E_{\mathrm{dif}}$ can be ignored, giving $E=E_{\mathrm{ph}}$ at any given moment of time. It is easy to show that the same result, $E=E_{{\mathrm{ph}}}$, also applies to the energy averaged over time $\tau$ \cite{next}.

The phonon energy of a liquid in the regime $\tau\gg\tau_{\rm D}$ is given, to a very good approximation, by the phonon energy of its solid. This is supported by the explicit equation for the liquid energy in the next section, and can be qualitatively discussed as follows. The only difference between the phonon states in a liquid and a solid is that the former does not support all transverse modes as a solid does, but only modes with frequency $\omega>\frac{1}{\tau}$ \cite{frenkel}. When $\tau\gg\tau_{\rm D}$, the fraction of missing transverse modes in a liquid is negligible and, furthermore, contributes a vanishingly small term to the phonon energy because the phonon density of states is proportional to $\omega^2$.

We therefore conclude that the energy of the viscous liquid is equal to the phonon energy, as in the solid. Consequently, Eq. (\ref{10}), derived for solids on the basis of phonons and Gr\"{u}neisen approximation, applies to viscous liquids too. This explains our earlier finding that the increase of liquid $c_v$ in the low-temperature viscous regime in Fig. 2a is well described by our proposed Eq. (\ref{10}).

\section{The origin of non-monotonic behavior of liquid $c_v$}

It is interesting to note the non-monotonic behavior of $c_v$ in Fig. 2 with a maximum. We explain this behavior as a result of two competing effects. On one hand, $c_v$ increases in the viscous regime due to anharmonicity as discussed above. On the other hand, $c_v$ decreases at high temperature as a result of progressively decreasing number of transverse waves with frequency $\omega>\frac{1}{\tau}$. We have studied this effect in a series of recent papers \cite{lenergy1,lenergy2,lenergy3}, and shown that the associated decrease of $c_v$ is in quantitative agreement with experimental data of many liquids. Explicitly, the energy of a classical liquid is \cite{lenergy2}:

\begin{equation}
E=NT\left(1+\frac{\alpha T}{2}\right)\left(3-\left(\frac{\tau_{\rm D}}{\tau}\right)^3\right)
\label{energy}
\end{equation}

At low temperature when $\tau\gg\tau_{\rm D}$, Eq. (\ref{energy}) gives $E=3NT\left(1+\frac{\alpha T}{2}\right)$ and $C_v=3N(1+\alpha T)$, Eq. (4). This is the result we observe in Fig. 2a at low temperature. At high temperature when $\tau\rightarrow\tau_{\rm D}$, the last term in Eq. (\ref{energy}), $\left(3-\left(\frac{\tau_{\rm D}}{\tau}\right)^3\right)$, can not be ignored. Its decrease with temperature dominates over $T\left(1+\frac{\alpha T}{2}\right)$ because $\tau$ decreases with temperature exponentially or faster. The result is that in the low-viscous regime $\tau\rightarrow\tau_{\rm D}$, $c_v$ decreases with temperature \cite{lenergy1,lenergy2,lenergy3}. The combination of two competing effects gives the maximum of $c_v$ as is seen in Fig. 2a.

We finally note that experimentally, the non-monotonic behavior of $c_v$ shown in Figure 2a is challenging to observe. On one hand, the noticeable decrease of $c_v$ requires $\tau$ approaching $\tau_{\rm D}$ as discussed above and, therefore, requires experimenting with low-viscous liquids such as metallic, noble-atom and some molecular liquids \cite{lenergy3}. These liquids tend to easily crystallize on cooling, preventing the formation of the viscous regime and accompanied linear increase of $c_v$. On the other hand, the linear increase of $c_v$ could be observed in viscous liquids such as silicates, chalcogenide and other systems. However, reaching low-viscous regime $\tau\rightarrow\tau_{\rm D}$ in these systems and accompanied decrease of $c_v$ due to the loss of transverse waves requires high temperatures where experiments are challenging. Moreover, viscous liquids often have strong bonds and high Debye temperature of internal vibrations, with the result that $c_v$ continues to increase even at high temperature due to progressive excitation of internal vibrations, counteracting the decrease of $c_v$ due to the loss of transverse modes. As a result, experiments typically observe either decrease of $c_v$ in low-viscous liquids or increase of $c_v$ in high-viscous liquids but not both. These problems did not originate in our MD simulations in which we were able to reach both low-viscous ($\tau\rightarrow\tau_{\rm D}$) and high-viscous liquid state ($\tau\gg\tau_{\rm D}$) and which, furthermore, were classical.

\section{Summary}

In summary, we have discussed a new way of evaluating the effects of anharmonicity on system's thermodynamic functions such as heat capacity, and have demonstrated its good predictive power. Importantly, our theory can be used to evaluate anharmonic $c_v$ in a system of any complexity including glasses and viscous liquids, in contrast to previous treatments of anharmonicity. In liquids, anharmonicity results in the increase of $c_v$ at low temperature, contributing to a non-monotonic behavior and a maximum of $c_v$.


\begin{thebibliography}{99}
\bibitem{zir3} The parameters of Si--O Morse interaction for ZrSiO$_4$ interatomic potential reported in Ref. \cite{zir1} contained a typo. The correct parameters of the Morse interaction, $y(r)=D\left(\exp(-2k(r-r_0))-2\exp(-k(r-r_0))\right)$, are: $D$=1.252 eV, $r_0$=1.627 \AA\ and $k=$2.83 \AA$^{-1}$.
\end{thebibliography}
\end{document}